\documentclass{article}

\usepackage{amsfonts}
\usepackage{amstext}
\usepackage{color}
\usepackage{amsthm}
\theoremstyle{plain}
\definecolor{Red}{cmyk}{0,1,1,0}

\definecolor{Blue}{cmyk}{1,1,0,0}

\definecolor{Pink}{cmyk}{0,1,0,0}

\definecolor{Green}{cmyk}{1,0,1,0.5}

\newcommand{\ba}{\begin{array}}
\newcommand{\ea}{\end{array}}
\newcommand{\be}{\begin{equation}}
\newcommand{\ee}{\end{equation}}
\newcommand{\ben}{\begin{enumerate}}
\newcommand{\een}{\end{enumerate}}

\newcommand{\F}{ {\cal F} }

\newcommand{\tra}{\widehat}

\newcommand{\R}{\mathbb{R}}

\newcommand{\B}{{\cal B}}

\newcommand{\N}{\mathbb{N}}

\newtheorem{theorem}{Theorem}[section]
\newtheorem{lema}{Lemma}[section]

\title{
{\large{ \bf{ASYMPTOTICS FOR NONLINEAR INTEGRAL EQUATIONS WITH GENERALIZED HEAT KERNEL AND TIME DEPENDENT COEFFICIENTS USING RENORMALIZATION GROUP TECHNIQUE}}}}

\date{} 

\begin{document}
\maketitle

\centerline{\scshape Gast\~ao A. Braga}
\medskip
{\footnotesize
 \centerline{Departamento de Matem\'atica}
  \centerline{Universidade Federal de Minas Gerais}
   \centerline{Caixa Postal 1621, Belo Horizonte, 30161-970, Brazil}
} 

\medskip

\centerline{\scshape Jussara M. Moreira}
\medskip
{\footnotesize
 \centerline{Departamento de Matem\'atica}
  \centerline{Universidade Federal de Minas Gerais}
   \centerline{Caixa Postal 1621, Belo Horizonte, 30161-970, Brazil}
}

\medskip

\centerline{\scshape Camila F. Souza}
\medskip
{\footnotesize
 \centerline{Departamento de Matem\'atica}
  \centerline{Centro Federal de Educa\c c\~ao Tecnol\'ogica de Minas Gerais}
   \centerline{Av. Amazonas, 5.253, Nova Sui\c ca, 30421-169, Brazil}
}

\def\l{\lambda}

\baselineskip = 22pt

\maketitle

\begin{abstract}
In this paper we employ the Renormalization Group (RG) method
to study the long-time asymptotics of a class of nonlinear integral equations with a generalized heat kernel and with time-dependent coefficients.
The nonlinearities are classified and studied according to its role in the asymptotic behavior. Here we prove that adding nonlinear perturbations classified as irrelevant, the behavior of the solution in the limit $t \to\infty$ remains unchanged from the linear case. In a companion paper, we will include a type of nonlinearities called marginal and we will show that, in this case, the large time limit gains an extra logarithmic decay factor.
\end{abstract}
\clearpage

\section{\large{Introduction}}
\label{sec:intr}
The purpose of this paper is to obtain the long-time behavior of solutions to the
integral equation
$$
u(x,t)=\int_\R {G(x-y, s(t))f(y)dy}\,\, +
$$
\begin{equation}
\label{equ:nao:lin:int}
\int_1^{t}\int_\R {G(x-	y,s(t)-s(\tau))F(u(y,\tau))dy d\tau},~~ x\in\R\mbox{ and } t>1,
\end{equation}
using the Renormalization Group (RG) method as developed by Bricmont et al.
\cite{bib:bric-kupa-lin}. $G=G(x, t)$ is a generalized kernel satisfying the following hypotheses (which we denote by {{\bf (G)}}):
\begin {enumerate}
\item[$(i)$]
There are integers $q >1$ and $M>0$ such that $G(\cdot,1)\in C^{q+1}(\R)$ and
$$
\sup_{x\in\R}\{(1+|x|)^{M+2}|G^{(j)}(x, 1)|\}<\infty, ~~~~j=0,1,..., q +1,
$$
where $G^{(j)}(x, 1)$ denotes the j-th derivative $(\partial_x^jG)(x, 1)$.
\item[$(ii)$]
There is a positive constant $d$ such that
$$
G(x,t)=t^{-\frac{1}{d}}G\left(t^{-\frac{1}{d}}x,1\right),~~~~ x\in\R,~t>0;
$$
\item [$(iii)$]
$$
G(x,t)= \int_\R{G(x-y, t-s)G(y,s)dy}~~~ \mbox{ for } x\in\R \mbox{ and } t>s>0;
$$
\item [$(iv)$] $G(x, t)$ is not identically zero and $G(x, t)\geq 0$, for $x\in \R$ and $t>0$.
\end{enumerate}
The function $s(t)$ appearing in the $G(x,t)$ argument in equation (\ref{equ:nao:lin:int}) is
\begin{equation}
\label{def:s(t):int}
s(t)=\int_{1}^{t}c(\tau)d\tau = \frac{t^{p+1}-1}{p+1}+r(t),
\end{equation}
where $c(t)$ is a positive function in $L^1_{loc}((1,+\infty))$ of type
$t^p+\mbox{o}(t^p)$, with $p>0$ and $o(t^p)$ is a little order of $t^p$ as $t\rightarrow \infty$.
We assume that $f$ belongs to a certain Banach space that will be specified later and $F(u)=\sum_{j\geq \alpha}{a_ju^j}$, where $\alpha$
will be chosen according to $p$ and $d$.

By analogy with the initial value problem for an evolution equation we call (\ref{equ:nao:lin:int})
{\em{initial value problem (IVP)}} and $ f $ initial data, although, since the kernel $G(x,t)$ is not specified, the integral equation generalizes these types of problems. This approach was adopted in \cite{bib:Ish:Kawa:Koba:1, bib:Ish:Kawa:Koba} by K. Ishige, T. Kawakami and
K. Kobayashi. They proved that, with similar conditions as those imposed in $(i)$, $(ii)$, $(iii)$ above, the solution to
(\ref{equ:nao:lin:int}), with $s(t)=t$ and $\alpha > 1+d$ behaves as
$$
\frac{A}{t^{1/d}}G\left(\frac{x}{t^{1/d}}, 1\right)\mbox{  when  } t\rightarrow \infty.
$$
Here we extend this result proving that, if $s(t)$ is of type (\ref{def:s(t):int}), with $p>0$ and if $\alpha$ is an integer bigger than $(p+ 1+ d)/(p+1)$, then,
$$
u(x,t) \sim \frac{A}{t^{(p+1)/d}}G\left(\frac{x}{t^{(p+1)/d}},
\frac{1}{p+1}\right)\mbox{ when } t\rightarrow \infty,
$$
which is essentially the statement of Theorem \ref{teo:pri:cas:irr}.

Our proof relies on the Renormalization Group (RG) approach. The RG method
was originally introduced in quantum field theory \cite{bib:gell-low} and statistical mechanics \cite{bib:wilson1}
and it was afterwards applied to the asymptotic analysis of deterministic differential equations, both analytically (\cite{bib:gold-book}, \cite{bib:bric-kupa}, \cite{bib:bric-kupa-lin}) and numerically \cite{bib:chen-gold}. It proved to be very useful on the asymptotic analysis in problems involving an infinite number of scales.
In equation (\ref{equ:nao:lin:int}), the multiple scales refer to rewrite the equation formulated
for $ t> 1$ as an infinite superposition of equations formulated for  $ t \in [1, L] $, with $ L> 1$. Our result here is a generalization of the problem presented in \cite{bib:braga-furt-mor-rolla-tp}, where the Renormalization Group method was applied to study the asymptotic behavior of the solution to I.V.P. $u_t  =c(t)u_{xx}+\lambda F(u)$, $t >1$, $x\in \R$, $u(x,1)  =  f(x)$ with $c(t)=t^p+o(t^p)$ and nonlinearity of type
$F(u)=\sum_{j\geq \alpha }{a_ju^j}$ with $\alpha$ integer greater than $(p+3)/(p+1)$, which is a particular case of those obtained in this paper.

In order to state the main theorem which will be proved in here, we first define the Banach space for the initial data $f$.
Let $q>1$ be given in $(i)$ of $\textbf{(G)}$, then
$$
\B_q \equiv \{f:\R\rightarrow\R~|~\tra f(\omega)\in C^1(\R)\mbox{ and }\|f\|<\infty\},
$$
with $\|f\|=\sup(1+|\omega|^q)(|\tra f(\omega)|+|\tra{f}'(\omega)|)$.
Finally, consider the integral equation (\ref{equ:nao:lin:int}) under the following hypothesis:
\begin{enumerate}
\item[$\textbf{(I)}$]
\begin{enumerate}
\item[$(I_1)$]  $f\in\B_q$;
\item[$(I_2)$] $c(t)$ is a positive function in $L^1_{loc}((1,+\infty))$ and $c(t) = t^p + o(t^p)$ as $t\to\infty$, with $p > 0$;
\item[$(I_3)$]$F(u)=\lambda\sum_{j\geq\alpha}{a_ju^j}$ analytic at $u = 0$, with convergence radius $\rho > 0$, $\alpha$ integer such that
$\alpha>(p+1+d)/(p+1)$ and $\lambda\in[-1,1]$.
\end{enumerate}
\end{enumerate}
\begin{theorem}
\label{teo:pri:cas:irr}
Consider equation (\ref{equ:nao:lin:int}) under the hypothesis $\textbf{(I)}$. Then, there exists $\epsilon>0$ such that, if $\|f\|<\epsilon$, then
(\ref{equ:nao:lin:int}) has a unique solution $u$ which satisfies, for certain $A=A(d,f,F,p)$,
\begin{equation}
\label{eq:pri:cas:irr}
\lim_{t\rightarrow\infty}\|t^{(p+1)/d}u(t^{(p+1)/d}\cdot,t)-Af_p^*\|=0
\end{equation}
with
\begin{equation}
\label{def:f_p^*}
f_p^*(x)=G\left(x, \frac{1}{p+1}\right).
\end{equation}
\end{theorem}

This paper proceeds as follows. In Section \ref{sec:linear} we employ the RG approach to
the linear equation (\ref{equ:nao:lin:int}), with $F\equiv 0$ in order to establish
how the method works with integral equations and to obtain results which will be
useful in the nonlinear case developed in Section \ref{sec:irrelevant}.

\section{\large{The linear case}}
\label{sec:linear}

In this section we present some properties of the kernel $G$ and some results concerning the employment of the RG method to $u(x,t)$ given by the linear piece of (\ref{equ:nao:lin:int}):
\begin{equation}
\label{equ:int}
u(x,t)=\int{G(x-y, s(t))f(y)dy},
\end{equation}
for $t>1$, $x\in\R$, $G=G(x, t)$ and $s(t)$ satisfying, respectively, hypotheses $\textbf{(G)}$ and equation (\ref{def:s(t):int}) and with
$f \in \B_q$.
The results obtained in this case will guide us throughout the studies of the nonlinear equation in the next sections.
We will prove the following theorem:
\begin{theorem}
\label{Teo:pri} Let $G$ be a kernel satisfying conditions {\bf (G)}, $f\in \B_{q}$, $A=\tra f(0)$ and $f_p^*(x)$ given by (\ref{def:f_p^*}) and
consider $u(x,t)$ given by (\ref{equ:int}). Then,
\begin{equation}
\label{lim:pri}
\lim_{t\rightarrow \infty}\|t^\frac{p+1}{d}u(t^\frac{p+1}{d}.,t)-Af^*_p(.)\|=0.
\end{equation}
\end{theorem}
Rather than proving the above theorem, our greatest interest is to establish the method of the renormalization group applied to an integral equation.
In the RG approach the long-time behavior of solutions to PDEs is related to the
existence and stability of fixed points of an appropriate RG transformation. Once a proper RG transformation has been found for a particular problem,
the method is iterative and the application of the RG transformation progressively evolves the solution in time and
at the same time renormalizes the terms of the equation.
In order to define the RG operator for problem (\ref{equ:int}), let $L>1$ be given and define 
$$
u_0(x,t)=\int{G(x-y, t-1)f(y)dy}, \,\,\, t\in(1,L],
$$
$$
u_{n}(x,t) \equiv \int{G(x-y,s(t)-s(L^n))u_{n-1}(y, L^n)dy}, \,\,\, t\in(L^n,L^{n+1}], \,\,\, n=1, 2, \cdots,
$$
and, for $n=0, 1, 2, \cdots$,
\begin{equation}
\label{def:ufn}
u_{f_n}(x,t)\equiv\int{G\left(x-y,\frac{s(L^nt)-s(L^n)}{L^{n(p+1)}}\right) f_n(y)}dy, \,\,\, t\in(1,L],
\end{equation}
with $f_0 \equiv f$ and
\begin{equation}
\label{def:rg:lin}
f_{n+1}\equiv R^0_{L,n}f_n \equiv L^{(p+1)/d}u_{f_n}(L^{(p+1)/d}\cdot,L).
\end{equation}
The nth step RG operator for the linear equation (\ref{equ:int}) is defined by (\ref{def:rg:lin}) and it is not hard to see that
\begin{equation}
\label{eq:prop-semi-grup}
R^0_{L^n,0}f_0=L^{n(p+1)/d}u(L^{n(p+1)/d}\cdot,L^n)= (R^0_{L,n-1}\circ...\circ R^0_{L,1}\circ R^0_{L,0})f_0
\end{equation}
and therefore the limit $\lim_{t\rightarrow\infty}t^{(p+1)/d}u(t^{(p+1)/d}x,t)$ is equivalent to taking the $ n \to \infty $ limit on
the right hand side of the above equality. Our goal is to study the dynamics of $ R^0_{L,n}f_n$, $n = 0, 1, 2, \cdots$,
in the space of the initial data.

The method involves a decomposition of the initial data into two terms: one in the direction of
$R^0_{L,n}f^*_p$ and the other which will be contracted by the operator. In fact, one of the basic results that will guarantee the success
of the method is the \textit{Contraction Lemma} \ref{lemadacontracao}, which establishes basically that, for $L$ large enough,
the operator $R^0_{L, n}$ is a contraction in the space of functions $g\in \B_q$ such that $\hat g(0) = 0$.
This result and Lemma \ref{lem:pon:fix}, which assures that $f_p^*$ is an asymptotic fixed point of the RG operator,
combine to prove that the function $u (x, t)$, defined by (\ref{equ:int}) and duly rescaled, behaves asymptotically as a
multiple of $f_p^*$, proving Theorem \ref{Teo:pri}.

\subsection{Preliminaries}
In order to prove Theorem \ref{Teo:pri} we will establish some properties of the kernel $G$ which follow from conditions {{\bf (G)}}:
\begin{lema}
\label{Lem:Lim:G}
If $G(x,t):\R\times(0,\infty)\to \R$ satisfies property $(i)$ of $\textbf{(G)}$, then $\tra G(\omega,1), \tra G'(\omega,1)\in L^\infty(\R)$ are well defined,
$\sup_{\omega\in\R}(1+|\omega|^q)|\omega||\tra G(\omega, 1)|<\infty$ and $\sup_{\omega\in\R}(1+|\omega|^q)|\omega||\tra G'(\omega, 1)|<\infty$.
\end{lema}

{\bf{Proof: }} It follows from property (i) of {{\bf (G)}} and from Fourier Transform results. \qed

From the above lemma we can define
\begin{equation}
\label{cons:cond:1.1:G}
K \equiv \sup_{\omega\in\R}|\tra G(\omega,1)|
\end{equation}
and
\begin{equation}
\label{cons:cond:1.1:G'}
K_1 \equiv \sup_{\omega\in\R}|\tra G'(\omega,1)|
\end{equation}
and also $\sup_{\omega\in\R}(1+|\omega|^q)|\tra G(\omega, 1)|<\infty$ and $\sup_{\omega\in\R}(1+|\omega|^q)|\tra G'(\omega, 1)|<\infty$.
It follows also from Lemma \ref{Lem:Lim:G} and $(ii)$ of $\textbf{(G)}$ that
$\tra G(\omega,t)$ is well defined for all $t>0$ and we can rewrite condition $(ii)$ in the Fourier Space as
\begin{equation}
\label{cond:1.2:tra G}
\tra G(\omega,t)=\tra G(t^{\frac{1}{d}}\omega,1),~~~~\mbox{ for } t>0 \mbox{ and } \omega\in\R
\end{equation}
and also obtain
\begin{equation}
\label{cond:1.2:tra G'}
\tra G'(\omega,t)=t^{\frac{1}{d}}\tra G'(t^{\frac{1}{d}}\omega,1),
~~~~\mbox{ for } t>0 \mbox{ and } \omega\in\R.
\end{equation}
Finally, condition $(iii)$ of $\textbf{(G)}$ implies that
\begin{equation}
\label{cond:1.3:tra G}
\tra G(\omega,t)=\tra G(\omega,t-s)\tra G(\omega, s) ~~ t>s>0 \mbox{ and } \omega\in\R.
\end{equation}
These results together lead to the following Lemma:
\begin{lema}
\label{G: decres}
Suppose $G(x,t):\R\times(0,\infty)\to \R$ satisfies properties $(i)$, $(ii)$ and $(iii)$ of $\textbf{(G)}$.
Given $t_1, t_2 \in (0,\infty)$, with $t_1<t_2$, then, for all $\omega\in\R$,
$$
|\tra G(\omega,t_2)|\leq K|\tra G(\omega,t_1)|
$$
and
$$|\tra G'(\omega,t_2)|\leq K_1(t_2-t_1)^\frac{1}{d}|\tra G(\omega,t_1)|+K|\tra G'(\omega,t_1)|,
$$
with $K$ and $K_1$ given by (\ref{cons:cond:1.1:G}) and (\ref{cons:cond:1.1:G'}), respectively.
\end{lema}

\begin{lema}
\label{lem:tra G(0,t)}
If $G(x,t):\R\times(0,\infty)\to \R$ satisfies conditions $(i)$ to $(iv)$ of $\textbf{(G)}$, then, for all $t>0$,
$$\int_{\R} G(x,t)dx=1.
$$
\end{lema}
{\bf{Proof:}} Condition $(i)$ guarantees the validity of Lemma \ref{Lem:Lim:G} which, together with conditions $(ii)$ and $(iii)$ lead to
(\ref{cond:1.2:tra G}) and (\ref{cond:1.3:tra G}). Condition $(iv)$ implies that $\tra G(0,t) >0$, for $t>0$. Then, from (\ref{cond:1.3:tra G}) with $s=1$
 and $t=2$, we have $\tra G(0,2) = \tra G(0,1)\tra G(0,1)$ and from (\ref{cond:1.2:tra G}), $\tra G(0,t)=\tra G(0,1)$ for all $t>0$.
 Therefore, $\tra G(0,1)=\tra G(0,2) = [\tra G(0,1)]^2$ which proves the lemma, since $\tra G(0,1) >0$.
\qed

We now enunciate and prove some results and properties of the RG operator
defined in (\ref{def:rg:lin}) that will be used to obtain the asymptotic behavior, both for the linear and nonlinear cases. To simplify the notation, from now on we denote $R^0_{L,0}f\equiv R^0_{L}f$.

\begin{lema}
\label{lem:pon:fix}
There is $L_1>1$ and positive constants $\tilde{K}$, $M=M(p,q,d)$ and $C_{d,p,q}$, such that, for $L>L_1$ given,
\begin{equation}
\label{cot:f_p^*}
\|f_p^*\|< C_{d,p,q}.
\end{equation}
\begin{equation}
\label{cot:f^*_p}
\|R^0_{L^n}f^*_p\|\leq \tilde{K}
\end{equation}
and
\begin{equation}
\label{equ:pon:fix}
\|R_{L^n}^0f^*_p-f^*_p\|\leq M\left|\frac{r(L^n)}{L^{n(p+1)}}\right|^{\frac{1}{d}},
\end{equation}
where $f^*_p(x)$ and $r(t)$ are defined, respectively by (\ref{def:f_p^*}) and (\ref{def:s(t):int}).
\end{lema}

{\bf{Proof:}} Using the definition of the $\B_q$ norm and the properties of $G$ we obtain (\ref{cot:f_p^*}), with
$$
C_{d,p,q}=(p+1)^{\frac{{q}}{d}}\sup_{k \in \R}(1+|k|^{q})[|\tra G(k,1)|+|\tra G'(k,1)|].
$$
Now, since, for a given $g\in \B_q$,
\begin{equation}
\label{equ:tra:R}
\F(R^0_{L,n}g)(\omega)=\tra G\Bigg(\frac{\omega}{L^{(p+1)/d}},s_n(L)\Bigg)\tra g\Bigg(\frac{\omega}{L^{(p+1)/d}}\Bigg)
\end{equation}
and $f_p^*(x)=G\left(x, \frac{1}{p+1}\right)$, using (\ref{cond:1.3:tra G}) it is not hard to see that $\|R_{L^n}^0f^*_p\|$ is given by
{\small{
$$
\sup_{\omega \in \R}(1+|\omega|^{\tilde{q}})\Bigg[\Bigg|\tra G\left(\omega,
\frac{s_0(L^n)}{L^{n(p+1)}}+\frac{1}{(p+1)L^{n(p+1)}}\right)\Bigg|+
\Bigg|\tra G'\left(\omega,\frac{s_0(L^n)}{L^{n(p+1)}}+\frac{1}{(p+1)L^{n(p+1)}} \right)\Bigg|\Bigg].
$$
}}
Furthermore, there is $L_1>1$ such that, if $L>L_1$, then
\begin{equation}
\label{cot:s_n(L)}
\frac{1}{6(p+1)}<\frac{s_n(L)}{L^{p+1}}<\frac{3}{2(p+1)},
\end{equation}
for all $n\geq 0$. Therefore,
$$
\frac{1}{6(p+1)}<\frac{s_0(L^n)}{L^{n(p+1)}}+\frac{1}{(p+1)L^{n(p+1)}}<\frac{5}{2(p+1)},
$$
for $L>L_1$ and using Lemma \ref{G: decres} we obtain
(\ref{cot:f^*_p}) with
$$
\tilde{K}=[6(p+1)]^{\frac{{q}}{d}}\left[K+7K_1/3(p+1)\right]\sup_{k \in \R}(1+|k|^q)
[|\tra G\left(k,1\right)|+|\tra G'\left(k,1\right)|].
$$
Using (\ref{cond:1.2:tra G}) and (\ref{cond:1.3:tra G}) we get
\begin{equation}
\label{equ:f}
\F[R_{L^n}^0f^*_p-f^*_p](\omega)=\tra G\left(\omega,\frac{1}{p+1}\right)\left[\tra G\left(\left(\frac{r(L^n)}{L^{n(p+1)}}\right)^{\frac{1}{d}}\omega,1\right)-1\right]
\end{equation}
and using Lemma \ref{lem:tra G(0,t)} and the Mean Value Theorem we conclude that
\begin{equation}
\label{cot:G}
\left|\tra G\left(\omega\left(\frac{r(L^n)}{L^{n(p+1)}}\right)^{\frac{1}{d}},1\right)-1\right|\leq K_1\left|\frac{r(L^n)}{L^{n(p+1)}}\right|^{\frac{1}{d}}|\omega|.
\end{equation}
Deriving (\ref{equ:f}) and using (\ref{cot:G}) and (\ref{cons:cond:1.1:G'}) we get
$$
|[\F(R_{L^n}^0f^*_p-f^*_p)]'(\omega)|\leq K_1\left|\frac{r(L^n)}{L^{n(p+1)}}\right|^{\frac{1}{d}}\left[|\omega|\left|\tra G'\left(\omega,\frac{1}{p+1}\right)\right|+\left|\tra G\left(\omega,\frac{1}{p+1}\right)\right|\right]
$$
and combining (\ref{equ:f}) and the above bound we get
(\ref{equ:pon:fix}) with
$$
M=K_1(p+1)^{\frac{q}{d}}[1+(p+1)^{\frac{1}{d}}]\sup_{w \in \R}(1+|w|^q)[(|w|+1)|\tra G(w,1)|+|w||\tra G'(w,1)|],
$$
which is finit from Lemma \ref{Lem:Lim:G}.
\qed

\begin{lema}(Contraction Lemma)
\label{lemadacontracao}
There exist constants $C=C(d,p,q)>0$ and $L_1>1$ such that
\begin{equation}
\label{lema:contr}
\|R^0_{L,n}g\|\leq \frac{C}{L^{(p+1)/d}}\|g\|,~~\forall~L>L_1 {\mbox{ and }} n=0,1,2,\cdots,
\end{equation}
if $g\in \B_q$ is such that $\tra g(0)=0$.
\end{lema}
\textbf{Proof:}
Using (\ref{cond:1.2:tra G}), (\ref{cond:1.2:tra G'}) and (\ref{equ:tra:R}) we get
$$
\|{R^0_{L,n}(g)}\|\leq\sup_{\omega \in \R}(1+|\omega|^{q})\Bigg\{\Bigg[\Bigg| \tra G\left(\omega,\frac{s_n(L)}{L^{{p+1}}}\right)\Bigg|
+   \Bigg| \tra G'\left(\omega,\frac{s_n(L)}{L^{{p+1}}}\right)\Bigg|\Bigg]\Bigg|\tra g\Bigg(\frac{\omega}{L^{(p+1)/d}}\Bigg)\Bigg|
$$
$$  +\frac{1}{L^{(p+1)/d}}\Bigg| \tra G\left(\omega,\frac{s_n(L)}{L^{{p+1}}}\right)\Bigg|\Bigg| g'\Bigg(\frac{\omega}{L^{(p+1)/d}}\Bigg)\Bigg|\Bigg\}.
$$
Since $|\tra g'(\omega)|\leq \|g\|$ for all $\omega\in\R$ and $\tra g(0)=0$, we have $\left|\tra g\left(\frac{\omega}{L^{(p+1)/d}}\right)\right|\leq |\omega|L^{-(p+1)/d}\|g\|$. Using this bound, (\ref{cot:s_n(L)}) and Lemma \ref{G: decres}
we get as an upper bound for $\|{R^0_{L,n}g}\|$:
$$
\sup_{\omega \in \R}(1+|\omega|^{q})\Bigg\{\Bigg[\left(K+K_1\left(\frac{4}{3(p+1)}\right)^{\frac{1}{d}}\right)\Bigg| \tra G\left(\omega,\frac{1}{6(p+1)}\right)\Bigg|+
$$
$$
K\Bigg| \tra G'\left(\omega,\frac{1}{6(p+1)}\right)\Bigg|\Bigg]\frac{|\omega|}{L^{(p+1)/d}}\|g\|
 +\frac{K}{L^{(p+1)/d}}\Bigg| \tra G\left(\omega,\frac{1}{6(p+1)}\right)\Bigg|\|g\|\Bigg\}.
$$
Again, using (\ref{cond:1.2:tra G}), (\ref{cond:1.2:tra G'}) and a change of variables $k=1/[6(p+1)]^\frac{1}{d}\omega$ we obtain (\ref{lema:contr}),
with
\begin{equation}
\label{def:C(d,p,q)}
C=A(d,p,q)\sup_{k\in\R}\left(
(1+|k|^{q})[|k||\tra G\left(k,1\right)+|k||\tra G'(k,1)|+|\tra G(k,1)|]
\right)
\end{equation}
and
$$
A(d,p,q)=[6(p+1)]^{(1+q)/d}\{K[2+6(p+1)]^{1/d}]+K_18^{1/d}\}.
$$
\qed

\subsection{Proof of Theorem \ref{Teo:pri}}

In order to prove Theorem \ref{Teo:pri}, we first decompose the initial data as $\tra f_0(0)f_p^*+g_0$ and prove that each initial data for the renormalized problems has a similar decomposition $f_n=AR^0_{L^n}f_p^*+ g_n$, with $\tra g_n(0)=0$.
We will see that Theorem \ref{Teo:pri} will follow from this result, together with Lemmas \ref{lem:pon:fix}
and \ref{lemadacontracao}.
\begin{lema}
\label{cot:g_n}
Given $f\in \B_q$, define $A\equiv \tra f(0)$, $f_0\equiv f$ and let $f_n=R^0_{L,n-1}f_{n-1}$,
$n=1, 2, \cdots$. Given $L>L_1$, there are functions $g_n\in \B_q$, $n=0,1,2,...$, such that $\tra g_n(0)=0$ $\forall \,\, n$
and
\begin{equation}
\label{def:g_n}
f_0=Af_p^*+g_0,~~~~~~~~~~f_n=AR^0_{L^n}f_p^*+ g_n.
\end{equation}
Furthermore,
\begin{equation}
\label{nor:g_n}
\|g_n\|\leq \left(\frac{C}{L^{(p+1)/d}}\right)^n\|g_0\|,\,\,\, n=0, 1, 2, \cdots
\end{equation}
with $C=C(d,p,q)$ given by (\ref{def:C(d,p,q)}).
\end{lema}
\textbf{Proof: } The Lemma follows inductively from Lemma \ref{lemadacontracao}.
\qed

\emph{{Proof of Theorem \ref{Teo:pri}: }} Define $L_2\equiv \max\{L_1,C^{d/(p+1)}\}$
and, for $L>L_2$, since $f_n=R^0_{L^n}f$, using (\ref{def:g_n}) and (\ref{nor:g_n}) we get
$$
\|R^0_{L^n}f-Af^*_p\|\leq \left(\frac{C}{L^{(p+1)/d}}\right)^n\|g_0\|+|A|\|R^0_{L^n}f^*_p-f^*_p\|, ~~~~\forall n~\in \N.
$$
Since $R^0_{L^n}f(x)=L^{n(p+1)/d}u(L^{n(p+1)/d}x,L^n)$, from (\ref{equ:pon:fix}) it follows that
$$
\|L^{n(p+1)/d}u(L^{n(p+1)/d}\cdot ,L^n)-Af^*_p
\|\leq \left(\frac{C}{L^{(p+1)/d}}\right)^n\|g_0\|+M|A|\Big|\frac{r(L^n)}{L^{n(p+1)}}\Big|^\frac{1}{d}
$$
and, since $r(t)=o(t^{p+1})$, the above inequality gives the limit (\ref{lim:pri}) for $t_n=L^n$.
Given $\delta \in (0,1)$, take $L_3>L_2$ such that $L_3^{\delta(p+1)/d}>C$. Then, if $L>L_3$,
$$
\left(\frac{C}{L^{(p+1)/d}}\right)^n=\left(\frac{C}{L^{\delta(p+1)/d}}\right)^n\frac{1}{L^{n(p+1)(1-\delta)/d}}\leq \frac{1}{L^{n(p+1)(1-\delta)/d}}
$$
and, if $t=L^n$,
\begin{equation}
\label{nor:final2}
\|t^{(p+1)/d}u(t^{(p+1)/d}.,t)-Af^*_p\|\leq\frac{\|g_0\|}{t^{(p+1)(1-\delta)/d}}+M|A|\left|\frac{r(t)}{t^{p+1}}\right|^{1/d}.
\end{equation}
Estimate (\ref{nor:final2}) is also valid if we take $t =\tau L^n$, with $\tau \in [1,L]$ and $L > L_3$, which completes the proof for all $t > L_3$.
\qed

\section{\large{The nonlinear case}}
\label{sec:irrelevant}

Our aim in this section is to establish Theorem \ref{teo:pri:cas:irr}. In order to do that we consider
the integral equation
{\small{
\begin{equation}
\label{equacao:nao:lin}
u(x,t)=\int {G(x-y, s(t))f(y)dy}+\lambda\int_1^{t}\int {G(x-y,s(t)-s(\tau))F(u(y,\tau))dy d\tau},
\end{equation}
}}
defined for $t>1$ and $x\in\R$, with kernel $G(x, t)$ satisfying conditions {\bf (G)}, $s(t)$ given by (\ref{def:s(t):int}),
$f\in \B_q$ 
and $F(u)=\sum_{j\geq \alpha}a_ju^j$ analytic at $u = 0$ with $\alpha\geq 2$ integer. Also, without
loss of generality, we assume $\lambda \in [-1, 1]$ so that the estimates obtained will be valid uniformly with respect to $\lambda$.
We shall prove that, if condition $\textbf{(I)}$ is satisfied and for small initial data, the above equation has a unique solution which, dully rescaled, converges, as in the linear case, to a multiple of $f^*_p$ when $t\rightarrow\infty$.

\subsection{Change of Scales and Renormalization}

Assuming that the solution $u$ to the integral equation (\ref{equacao:nao:lin}) is globally well defined, we fix $L>1$ and consider the sequence
$\{u_n\}_{n=0}^{\infty}$ by
\begin{equation}
\label{def:un:cas:irr}
u_n(x,t)\equiv L^{n(p+1)/d}u(L^{n(p+1)/d}x,L^nt), ~~ t\in [1, L], ~~ n=0, 1, 2, \cdots.
\end{equation}
We first have to determine the integral equation to be satisfied by $u_n(x,t)$. Unlike the case of a partial differential equation,
where the rescheduling is sufficient to establish the renormalized equation to be satisfied by $u_n$ (see \cite{bib:bric-kupa-lin}), in the case of an integral
equation we have to explore the properties of the kernel in order to determine this equation, which we
will call \textit{the integral renormalized equation}.

It follows from (\ref{equacao:nao:lin}) and (\ref{def:un:cas:irr}) that $u_n(x,t)$ can be written as $u_n(x,t)=a(x,t)+b(x,t)$ where
$$
a(x,t)\equiv L^{n(p+1)/d}\int {G(L^{n(p+1)/d}x-y, s(L^nt))f(y)dy}\,\,\, +
$$
$$
L^{n(p+1)/d}\lambda\int_1^{L^n}\int {G(L^{n(p+1)/d}x-y,s(L^nt)-s(\tau))F(u(y,\tau))dy d\tau}
$$
and
{\small{
$$
b(x,t) \equiv \lambda L^{n(p+1)/d}\int_{L^n}^{L^nt}\int {G(L^{n(p+1)/d}x-y,s(L^nt)-s(\tau))F(u(y,\tau))dy d\tau}.
$$
}}
From conditions $(ii)$ and $(iii)$ of $\textbf{(G)}$ and Fubini's theorem, we get
$$
a(x,t)=L^{n(p+1)/d}\int G(x-\omega, s_n(t))\Big[\int {G(L^{n(p+1)/d}\omega-y,s(L^n))f(y)dy}+
$$
$$
\lambda \int_1^{L^n}{\int{G(L^{n(p+1)/d}\omega-y,s(L^n)-s(\tau))F(u(y,\tau))dy d\tau}\Big]d\omega},
$$
with
\begin{equation}
\label{def:s_n(t)}
s_n(t)\equiv\frac{t^{p+1}-1}{p+1}+r_n(t),
\end{equation}
where
\begin{equation}
\label{def:r_n}
r_n(t)=\frac{r(L^nt)-r(L^n)}{L^{n(p+1)}}
\end{equation}
and $r(t)$ is given by (\ref{def:s(t):int}).
Furthermore, defining \textit{the renormalized initial data} $f_n$ by
\begin{equation}
\label{def:fn}
f_n(x)\equiv L^{n(p+1)/d}u(L^{n(p+1)/d}x,L^n),
\end{equation}
we obtain
\begin{equation}
\label{equ:a(x,t)}
a(x,t)=\int{G(x-\omega,s_n(t))f_n(\omega)d\omega}.
\end{equation}
To develop the $b(x,t)$ parcel we use a change of variables
$y=L^{n(p+1)/d}\omega$ and $\tau=L^nq$ and apply property $(ii)$ of $\textbf{(G)}$ to get
\begin{equation}
\label{equ:b(x,t)}
b(x,t)=\lambda_n\int_1^{t}\int {G(x-\omega,s_n(t)-s_n(q))F_{L,n}(u_n) d\omega dq},
\end{equation}
where
\begin{equation}
\label{def:lambdan}
\lambda_n= L^{n[-\alpha(p+1)+p+1+d]/d}\lambda
\end{equation}
and
\begin{equation}
\label{def:Fln}
F_{L,n}(u_n)=\sum_{j\geq\alpha}a_jL^{n(\alpha-j)(p+1)/d}u_n^j.
\end{equation}
From (\ref{equ:a(x,t)}) and (\ref{equ:b(x,t)}) we obtain therefore the integral renormalized equation for $u_n$:
\begin{eqnarray}
\label{equ:ren:nao:lin:fin}
u_{n}(x,t)&=& \int G(x-y,s_n(t))f_n(y)dy \nonumber \\
&+& \lambda_n\int_1^t\int G(x-y,s_n(t)-s_n(q))F_{L,n}(u_{n}(y,q))dydq.
\end{eqnarray}
In \cite{bib:bric-kupa-lin}, Bricmont et al. introduced a formal classification for perturbations os the heat equation based on
its behavior after a change of scales. We will now extend this classification for the integral equation (\ref{equacao:nao:lin}).
Consider (\ref{equacao:nao:lin}) with a nonlinearity of type $F(u)=u^{\alpha}$ and let $u_1(x,t)=L^{(p+1)/d}u(L^{(p+1)/d}x,Lt)$.
From the previous calculus we get equation (\ref{equ:ren:nao:lin:fin}) for $u_1$, with
$\lambda_1= L^{[-\alpha(p+1)+p+1+d]/d}\lambda$ and $F_{L,1}(u_1)=u_1^\alpha$. Therefore, defining $d_F\equiv -\alpha(p+1)+p+1+d$ we notice that,
if $d_F=0$, then the equation remains unchanged after the change of scales. This kind of
nonlinearity is classified as {\em marginal}. If $d_F <0$, since $ L> 1 $, we have $\lambda_1 <\lambda$ and
the nonlinear perturbation is classified as {\em irrelevant} and, on the other hand, if $d_F> 0 $, $F$ is
classified as {\em relevant}. This classification can be extended for nonlinearities of type $F(u)=\sum_{j\geq\alpha}a_ju^j$, with radius of convergence
$\rho>0$. Defining
\begin{equation}
\label{def:alp:cri}
\alpha_c=\frac{p+1+d}{p+1},
\end{equation}
$F$ is irrelevant if $\alpha>\alpha_c$, marginal if $\alpha=\alpha_c$ and relevant if $\alpha<\alpha_c$. 

In this section we study the asymptotic behavior of the solution to the integral equation (\ref{equacao:nao:lin}) with an irrelevant $F$, that is,
$F(u)=\sum_{j\geq\alpha}a_ju^j$, with $\alpha > \alpha_c$ integer.

\subsection{Local Existence and Uniqueness of Solution}

The basic idea of the RG method is to reduce the long-time asymptotics problem to the
analysis of a sequence of finite-time problems obtained by iterating an operator (the RG
operator) which, for a given $L>1$, we will define by
\begin{equation}
\label{def:rg:nao:lin}
(R_{L,n}f_n)(x) \equiv L^{(p+1)/d}u_{n}(L^{(p+1)/d}x,L),
\end{equation}
where $u_n(x,t)$ is the solution to (\ref{equ:ren:nao:lin:fin}). In order to (\ref{def:rg:nao:lin}) be well defined, we need to guarantee the existence and uniqueness of all renormalized problems (\ref{equ:ren:nao:lin:fin}). Notice also that, from (\ref{def:un:cas:irr}) and (\ref{def:fn}), we have
\begin{equation}
\label{def:f_n}
f_0 = f  { \mbox{ and } } f_{n+1}= R_{L,n}f_{n} \mbox{  for  } n = 0, 1, \cdots.
\end{equation}
In fact, the operator satisfies the {\em semi-group property}
$$
R_{L^n,0}f_0=L^{n(p+1)/d}u(L^{n(p+1)/d}\cdot,L^n)= [R_{L,n-1}\circ...\circ R_{L,1}\circ R_{L,0}]f_0
$$
and therefore, in order to obtain the limit (\ref{eq:pri:cas:irr}), we study the dynamics of the operators $R_{L,n}$,
$n=0, 1, 2, \cdots$ in the space of initial data $\B_q$. This argument is rigorous if we prove that each renormalized problem has a unique solution.
We will then show in Lemma \ref{exi:equ:ren} that, for each $n$, there exists $\epsilon_n>0$
such that, if $\|f_n\|<\epsilon_n$, then the integral renormalized equation (\ref{equ:ren:nao:lin:fin}) has a unique solution for $t \in [1,L]$.
The argument is analogous to the one presented in \cite{bib:braga-furt-mor-rolla-tp}. To state the Lemma, we define, for $L>1$ given, the Banach space
$B^{(L)}=\{u:\R\times[1,L]\rightarrow\R; u(\cdot,t) \in \B_q, ~\forall~t\in [1,L]\}$,
where $\|u\|_L=\sup_{t\in[1,L]}\|u(\cdot,t)\|$ and, if $u_{f_n}$ is the solution to (\ref{equ:ren:nao:lin:fin}) with $\lambda_n=0$, we define the ball
$B_{f_n}\equiv\{u_n\in B^{(L)}:\|u_n-u_{f_n}\|\leq \|f_n\|\}$, 
and the operator
$T_n(u_n)\equiv u_{f_n}+N_n(u_n)$,
where
\begin{equation}
\label{def:Nn(u)}
N_n(u_n)(x,t)=\lambda_n\int_1^{t}\int {G(x-y,s_n(t)-s_n(\tau))F_{L,n}(u_n(y,\tau))dy d\tau}
\end{equation}
with $s_n(t)$, $\lambda_n$ and $F_{L,n}(u_n)$ given, respectively, by (\ref{def:s_n(t)}),
(\ref{def:lambdan}) and (\ref{def:Fln}).

\begin{lema}
\label{exi:equ:ren} Given $n\in\N$ and $L>1$, there exists $\epsilon_n>0$ such that, if $\|f_n\|<\epsilon_n$, then the integral equation
(\ref{equ:ren:nao:lin:fin}) has a unique solution in $B_{f_n}$.
\end{lema}
\textbf{Proof:} Using that $L>1$, the definition of $s_n(t)$ and $F_{L,n}$ and the properties of the kernel $G$, we obtain the estimates
\begin{equation}
\label{cot:fin:N_n}
\|N_n(u_n)\|_L\leq C_n L^{n[-\alpha(p+1)+p+1+d]/d}\|f_n\|^2
\end{equation}
and
$$
\|N_n(u_n)-N_n(v_n)\|_L\leq C_n L^{n[-\alpha(p+1)+p+1+d]/d}\|f_n\|\|u_n-v_n\|,
$$
where
$$
C_n=[1+K+K_1(s_n(L))^\frac{1}{d}]^2[2K+(s_n(L))^{\frac{1}{d}}](L-1)S(\rho_0),
$$
where $K$ and $K_1$ are given respectively by (\ref{cons:cond:1.1:G}) and (\ref{cons:cond:1.1:G'}),
$S(z)=\sum_{j\geq\alpha}\left(\frac{C}{2\pi}\right)^{j-1}j|a_j|z^{j-2}$ and $\rho_0=2\pi\rho[(2^{q+1}+3)\int_{\R}{\frac{1}{1+|x|^q}dx}]^{-1}$.

In order to guarantee that $u_n$ is in the region of analyticity of $F_{L,n}$, we take $\|f_n\|<(1+K+K_1s_n(L)^\frac{1}{d})^{-1}\rho_0$.
Since $-\alpha(p+1)+p+1+d<0$ and $L>1$, defining
\begin{equation}
\label{def:epsilonn}
\epsilon_n\equiv\min\left\{\frac{1}{2C_n},\frac{\rho_0}{1+K+K_1s_n(L)^\frac{1}{d}}\right\},
\end{equation}
if $\|f_n\|<\epsilon_n$, then $\|N_n(u_n)\|_L<\|f_n\|$ and $\|N_n(u_n)-N_n(v_n)\|_L<1/2\|u_n-v_n\|_L$ for all $u_n$, $v_n\in B_{f_n}$, which proves that
the integral equation (\ref{equ:ren:nao:lin:fin}) has a unique solution $u_n(x,t)$ in $B_{f_n}$. This also implies that $f_{n+1}(x)=L^{\frac{p+1}{d}}u_n(L^{\frac{p+1}{d}}x, L)$ is well defined.
\qed

It follows from (\ref{cot:s_n(L)}) that, if $L>L_1$, the constants $C_n$ are uniformly superiorly bounded by
\begin{equation}
\label{def:N}
\tilde{C}=\left[1+K+K_1\left(\frac{3L^{p+1}}{2(p+1)}\right)^{\frac{1}{d}}\right]\left[2K+\left(\frac{3L^{p+1}}{2(p+1)}\right)^{\frac{1}{d}}\right](L-1)S{(\rho_0)}.
\end{equation}
Therefore, defining
\begin{equation}
\label{def:sigma}
\sigma=\min\left\{\frac{1}{2\tilde{C}},\rho_0\left[1+K+K_1\left(\frac{3L^{p+1}}{2(p+1)}\right)^{\frac{1}{d}}\right]^{-1}\right\},
\end{equation}
it is clear that $\sigma<\epsilon_n$ for all $n$, which is essential for our analysis since we want to be able to start with a sufficiently small initial data $f$
that guarantees a unique solution for each problem (\ref{equ:ren:nao:lin:fin}) and which will later ensure a unique global solution to (\ref{equacao:nao:lin}).

\subsection{Renormalization}

The Renormalization Group method consists in changing the calculation of the limit $t\to \infty$ of the solution of the integral equation (\ref{equacao:nao:lin})
by the analysis of the sequence of the initial data of the integral equations (\ref{equ:ren:nao:lin:fin}) and, to do that, we decompose the initial data (\ref{def:fn}) into two parcels, the first being a multiple of the function $R^0_{L^n} f_p^*$ (which according to Lemma \ref{lem:pon:fix}
converges to $f_p^*$ when $n\to \infty$) and the second which will be contracted in the process. 

We will now denote $\nu_{n}(x)\equiv N_n(u_n)(x,L)$ and, therefore, if we take $\|f_n\|<\epsilon_n$, Lemma \ref{exi:equ:ren} implies that (\ref{equ:ren:nao:lin:fin})
has a unique solution which, at time $t=L$, can be written as $u_{n}(x,L)=u_{f_n}(x,L)+\nu_{n}(x)$, with $u_{f_n}(x,t)$ the solution to the linear equation, given by (\ref{def:ufn}). It follows from (\ref{def:rg:lin}), (\ref{def:rg:nao:lin}) and (\ref{def:f_n}) that
\begin{equation}
\label{eq:repfn}
f_{n+1}(x)=(R_{L,n}f_{n})(x)
=R^0_{L,n}f_{n}(x) + L^{(p+1)/d}\nu_{n}(L^{(p+1)/d}x).
\end{equation}
For the next lemmas, we will refer to the constants $L_1$, $C_{d,p,q}$ and $\tilde{K}$ given in Lemma \ref{lem:pon:fix}, $C$ given in the Contraction Lemma \ref{lemadacontracao} and
$\tilde{C}$ given by (\ref{def:N}).
\begin{lema}
\label{lema:renorm}
Given $k\in \N$ and $L>L_1$, suppose $f_{n}$ is well defined by (\ref{def:fn}) for $n=0,1,...,k+1$.
Then, there are constants $A_{n+1}$ and functions $g_{n+1}\in\B_q$ such that, for $n=0,1,...,k$,
$\tra g_{n+1}(0)=0$,
\begin{equation}
\label{equ:1:lem:ren}
f_0=A_0f_p^*+g_0,~~~~~~~~~~~~f_{n+1}=A_{n+1}R^0_{L^{n+1}}f_p^*+g_{n+1},
\end{equation}
\begin{equation}
\label{cotaA}
|A_{n+1}-A_n|\leq \tilde{C}L^{n[-\alpha(p+1)+(p+d+1)]/d}\|f_n\|^2,
\end{equation}
\begin{equation}
\label{cota:g}
\|g_{n+1}\|\leq CL^{-(p+1)/d}\|g_{n}\| +\tilde{M}L^{n[-\alpha(p+1)+(p+d+1)]/d}\|f_{n}\|^{2},
\end{equation}
with $A_0=\tra f_0(0)$ and
\begin{equation}
\label{def:M}
\tilde{M}=(L^{q(p+1)/d}+\tilde{K})\tilde{C}.
\end{equation}
\end{lema}
\textbf{Proof: } Defining $g_0 \equiv f_0 -
A_0f_p^*$, it follows that $g_{0}\in B_q$ and, since $\tra f_p^*(0)=1$, $\hat{g}(0) = 0$. By hypothesis, $f_1$
is well defined by $R_{L,0}f_0$ and, using (\ref{eq:repfn}) and the decomposition for $f_0$,
$f_{1}=A_1R^0_{L}f^*_p+g_1$,
with $A_{1}\equiv A_{0}+\widehat{\nu}_{0}(0)$ and $g_1\equiv
R^0_Lg_0+L^{p+1/d}\nu_{0}(L^{(p+1)/d}\cdot)-\widehat{\nu}_{0}(0)R^0_Lf^*_p$. It follows from
(\ref{equ:tra:R}) that $\F(R^0_Lg_0)(0)=0$ and $\F(R^0_{L}f_p^*)(0)=1$ and therefore, $\tra g_1(0)=0$, which proves (\ref{equ:1:lem:ren}) for $n=0$.
Now suppose $f_{n}$ is well defined for $n=0,1,...k$.  Using (\ref{equ:1:lem:ren})
with $n=k-1$, (\ref{eq:repfn}) and (\ref{eq:prop-semi-grup}),
\begin{equation}
\label{1def:f_{j+1}}
f_{k+1}(x)=A_kR^0_{L^{k+1}}f_p^*(x)+R^0_{L,k}g_k(x)+L^{(p+1)/d}\nu_k(L^{(p+1)/d}x).
\end{equation}
Defining $A_{k+1}=A_k+\tra\nu_k(0)$ and
\begin{equation}
\label{def:g_{j+1}}
g_{k+1}(x)= R^0_{L,k}g_k(x)+L^{(p+1)/d}\nu_k(L^{(p+1)/d} x)-\tra\nu_k(0)R^0_{L^{k+1}}f_p^*(x),
\end{equation}
we rewrite (\ref{1def:f_{j+1}}) as $f_{k+1}=A_{k+1}R_{L^{k+1}}^0f_p^*+g_{k+1}$, with $\tra g_{k+1}(0)=0$,
which proves (\ref{equ:1:lem:ren}) for $n=k$.
Inequality (\ref{cotaA}) follows from (\ref{cot:fin:N_n}), since $ |A_{n+1}-A_n|=|\tra
\nu_n(0)|\leq\|\nu_n\|\leq \|N(u_n)\|_L$ and $C_n \leq \tilde C$, for all $n$.
Using (\ref{cot:fin:N_n}) and (\ref{cot:f^*_p}),
$$
\|\hat{\nu}_{n}(0)R^0_{L^{n}}f_p^*\|\leq \|\nu_{n}\|\|R^0_{L^n}f_p^*\|\leq \tilde{K}C_nL^{[n[-\alpha(p+1)+p+d+1]/d}\|f_{n}\|^{2}
$$
and
$$
\|L^{(p+1)/d}\nu_{n}( L^{(p+1)/d}.)\|_L\leq L^{q(p+1)/d}C_nL^{[n[-\alpha(p+1)+p+d+1]/d}\|f_n\|^2.
$$
Finally, since $\hat{g}_{n}(0)=0$, taking $L>L_{1}$ and using the Contraction Lemma \ref{lemadacontracao},
\begin{equation}
\label{des:g_1}
\|g_{n+1}\|\leq \frac{C}{L^{(p+1)/d}}\|g_{n}\| +E_{n}L^{[n(-\alpha(p+1)p+1+d)]/d}\|f_{n}\|^{2},
\end{equation}
with $E_n= (L^{q(p+1)/d}+\tilde{K})C_n$. Since $E_n \leq \tilde{M}$, for all $n$, we get (\ref{cota:g}).
\qed

Now we will show that we can start with a sufficiently small initial data $f_0$ such that all the renormalized problems (\ref{equ:ren:nao:lin:fin}) have a unique solution.
For $\alpha > \alpha_c$, let $\delta \in (0,1)$ such that
\begin{equation}
\label{cond:delta}
(1-\delta)(p+1) < \alpha(p+1)-(p+1+d)
\end{equation}
and define
\begin{equation}
\label{def:Ldelta}
L_{\delta}\equiv \max\{L_1,[2C(1+C_{d,p,q})]^{d/\delta(p+1)} \}
\end{equation}
and
\begin{equation}
\label{def:D}
D\equiv 1 + \tilde{K}\sum_{j=0}^{\infty}\frac{1}{L^{j(p+1)(1-\delta)/d}}.
\end{equation}
\begin{lema}
\label{lem:con:g:f}
Consider $\delta \in (0,1)$ satisfying (\ref{cond:delta}) and let $L>L_{\delta}$. Then, there is $\bar\epsilon > 0$ such that, if $\|f_{0}\|<\bar\epsilon$, then
$f_{n}$ and $g_n$ given respectively by (\ref{def:fn}) and (\ref{equ:1:lem:ren}) are well defined for all $n\geq 1$ and satisfy
\begin{equation}
\label{con:fn}
\|f_{n}\|\leq D\|f_0\|
\end{equation}
and
\begin{equation}
\label{equ:gn}
\|g_{n}\|\leq\frac{\|f_0\|}{L^{n(p+1)(1-\delta)/d}}.
\end{equation}
\end{lema}
\textbf{Proof: } Given $\delta \in (0,1)$ satisfying (\ref{cond:delta}) and $L>L_{\delta}$, define
\begin{equation}
\label{epsilonbarra}
\bar{\epsilon}=\min \left\{\frac{\sigma}{D},\frac{1}{2 \tilde{M}D^{2}L^{(1-\delta)(p+1)/d}}\right\}
\end{equation}
and assume $\|f_{0}\|<\bar\epsilon$. Inequalities (\ref{con:fn}) and (\ref{equ:gn}) follow inductively from Lemma \ref{lema:renorm}. Indeed,
since $D>1$, $\|f_0\|<\epsilon_0$ and from Lemma \ref{exi:equ:ren}, $f_1=R_{L,0}f_0$ is well defined. Inequality (\ref{cota:g}) for $n=0$
together with Lemma \ref{lem:pon:fix} and the definition of $g_0$ leads to
$$
\left\|g_{1}\right\|\leq\frac{1}{L^{(p+1)(1-\delta)/d}}\left[\frac{C(1+C_{d,p,q})}{L^{\delta(p+1)/d}}
+L^{(p+1)(1-\delta)/d}\tilde{M}\left\|f_{0}\right\|\right]\left\|f_{0}\right\|.
$$
From the hypothesis, the sum in the brackets is less then one and we get (\ref{equ:gn}) for $n=1$. Using Lemma \ref{lem:pon:fix}, inequalities (\ref{cotaA})
and (\ref{equ:gn}) with $n=1$ and the representation (\ref{equ:1:lem:ren}) of $f_1$,
\begin{equation}
\label{cot:f1}
\|f_{1}\|\leq\|g_{1}\|+\tilde{K}|A_{1}|\leq \left[\frac{1}{L^{(p+1)(1-\delta)/d}}+
\tilde{K}(1+\tilde{C}\|f_{0}\|)\right]\|f_{0}\|
\end{equation}
and since $\tilde{C}<\tilde{M}$, the definition of $\bar\epsilon$ and inequality (\ref{cot:f1}) imply that (\ref{con:fn}) is valid for $n=1$.
Now suppose that the Lemma is true for $n=k$. Since the hypothesis of Lemma \ref{lema:renorm} are satisfied, (\ref{cota:g}) holds for
$n=k$ and using the induction hypothesis,
$$
\|g_{k+1}\|\leq \frac{1}{L^{(k+1)(p+1)(1-\delta)/d}}\left[\frac{C}{L^{\delta(p+1)/d}}+
\frac{L^{k[-\alpha(p+1)+(p+d+1)]/d}}{L^{(p+1)(\delta-1)(k+1)/d}}\tilde{M}D^{2}\|f_0\|\right]\|f_0\|,
$$
which proves that (\ref{equ:gn}) holds for $n=k+1$. Furthermore, since $\|f_{k}\|\leq D\|f_0\|<\sigma<\epsilon_k$, from Lemma \ref{exi:equ:ren}, $f_{k+1}$
given by (\ref{def:f_n}) is well defined and, using representation (\ref{equ:1:lem:ren}) for $f_{k+1}$,
it follows from (\ref{equ:gn}) with $n=k+1$, Lemma \ref{lem:pon:fix} and $A_{k+1} = A_0+\sum^{k}_{j=0}(A_{j+1}-A_{j})$ that
$$
\|f_{k+1}\|\leq \frac{\|f_{0}\|}{L^{(1-\delta)(k+1)(p+1)/d}}+\tilde{K}\left(|A_0|+\sum^{k}_{j=0}|A_{j+1}-A_{j}|\right)
$$
and using (\ref{cotaA}) and (\ref{con:fn}) with $n=0,1,2,...k$, we get
$$
\|f_{k+1}\|\leq \left[\frac{1}{L^{(1-\delta)(k+1)/d}}+\tilde{K}\left(1+\tilde{C}D^2\|f_0\|\sum^{k}_{j=0}L^{j[-\alpha(p+1)+p+1+d]/d}\right)\right]\|f_{0}\|.
$$
Again, since $\tilde{C}<\tilde{M}$, the definition of $\bar\epsilon$ implies that (\ref{con:fn}) holds for $n=k+1$, ending the proof.
\qed

\subsection{Proof of Theorem \ref{teo:pri:cas:irr}}
\label{sec:prov-teor}

If $\|f_0\|<\bar\epsilon$ and $L>L_\delta$, it follows that the integral equation (\ref{equacao:nao:lin}) has a unique solution $u$ and, from Lemma \ref{lema:renorm},
 the semigroup property 
 and (\ref{equ:gn}),
$$\|L^{n(p+1)/d}u(L^{{n(p+1)/d}}.,L^{n})-A_nR^0_{L^n}f^*_p\|\leq L^{n(p+1)(\delta-1)/d}\|f_0\|.$$
From Lemmas \ref{lema:renorm} and \ref{lem:con:g:f}, we conclude that
\begin{equation}
\label{eq:dif-as}
|A_{n+1}-A_{n}|<
\frac{L^{n[-\alpha(p+1)+p+1+d]}}{2L^{(p+1)(1-\delta)/d}}\|f_0\|
\end{equation}
and, since $\alpha>(p+1+d)/(p+1) = \alpha_c$, it follows that $A_n \to A$ and we can bound $\|L^{n(p+1)/d}u(L^{n(p+1)/d}.,L^{n})-Af^*_p\|$ by
$$
|A|\|R^0_{L^n}f^*_p-f^*_p\|+\frac{\|f_0\|}{L^{n(p+1)(1-\delta)/d}}+|A_n-A|\|R^0_{L^n}f^*_p\|
$$
which, from Lemma \ref{lem:pon:fix}, goes to zero when $n\to\infty$. In fact, we can estimate the rate of convergency using (\ref{eq:dif-as})
and Lemma \ref{lem:pon:fix} to obtain the following upper bound for $\|L^{n(p+1)/d}u(L^{n(p+1)/d}.,L^{n})-Af^*_p\|$:
{\small{
$$
|A| \tilde{M}\Big|\frac{r(L^n)}{L^{n(p+1)}}\Big|^{\frac{1}{d}}+\frac{\|f_0\|}{L^{n(p+1)(1-\delta)/d}}+
\frac{L^{-n[\alpha(p+1)-(p+1+d)]/d}}{2L^{(p+1)(1-\delta)/d}(1-L^{[-\alpha(p+1)+p+1+d]/d})}K_{d,p,q}\|f_0\|,
$$
}}
which is valid for all $n>n_0$. To finish the proof, we proceed, as in the linear case, obtaining from the above bound, a similar inequality as (\ref{nor:final2}):
\begin{eqnarray*}
\|t^{(p+1)/d}u(t^{(p+1)/d}.,t)-Af^*_p(.)\|\leq &|A| \tilde{M}|t^{-(p+1)/d}r(t)|+t^{(p+1)(\delta-1)/d}\|f_0\|+ \\
&K_{d,p,q}\|f_0\|\frac{t^{[-\alpha(p+1)+(p+1+d)]/d}}{2{L_{\delta}}^{(p+1)(1-\delta)/d}(1-{L_\delta}^{[-\alpha(p+1)+p+1+d]/d})}.
\end{eqnarray*}
\qed


\end{document}